\documentclass{elsart1p}

\usepackage{graphicx}
\usepackage{amssymb}

\newcommand{\dd}{\mathrm{d}}

\begin{document}

\begin{frontmatter}




\title{Heavy-Quark Kinetics at RHIC and LHC}


\author[label1]{V. Greco\corauthref{cor1}},
\author[label2]{H. van Hees},
\author[label2]{R. Rapp}
\address[label1]{Dipartimento di Fisica e Astronomia, Via S. Sofia 64,
  I-95125 Catania, Italy}
\address[label2]{Cyclotron Institute and Physics Department, 
       Texas A\&M University, College Station, Texas 77843-3366, U.S.A.}
\corauth[cor1]{greco@lns.infn.it}

\begin{abstract}
  In ultrarelativistic nuclear collisions heavy quarks are produced out
  of thermal equilibrium in the very early stage of the reaction and
  their thermalization time was expected to be considerably larger than
  that of light quarks. On the other hand, a strongly-interacting QGP
  can be envisaged in the heavy quark sector due to the presence of
  heavy-light hadron-like resonances in the QGP for temperatures up to
  $\sim$$2 T_{C}$.  We investigate the consequences of such states
  employing a relativistic Langevin simulation. Hadronization is modeled
  by a coalescence-fragmentation scheme. We present the predictions for
  the nuclear modification factor and elliptic flow of $D$ and $B$
  mesons at LHC energies and compare the results with the successful
  predictions of this model for RHIC conditions. We find similar
  heavy-quark thermalization effects at LHC and RHIC.
\end{abstract}

\begin{keyword}
Heavy Quarks, Quark-Gluon Plasma, Collective flow, Hadronization.
\PACS 12.38 Mh, 24.85.+p, 25.75 Nq, 25.75 Ld
\end{keyword}
\end{frontmatter}

\section{Introduction}

For heavy quarks, their mass $m_Q$$\gg$$T_c$$\simeq$$180$~MeV (critical
temperature).  Hence their kinetic equilibration time is expected to be
larger than that for light partons, thus showing better sensitivity to
the in-medium interactions. Due to a slower thermalization a smaller
elliptic flow of $D$ mesons, $v_{2}^{D}$, was expected; however, it was
suggested that a sizable $v_{2}^{D}$ could result from that of the light
quarks~\cite{Greco:2003vf} through a coalescence hadronization
mechanism~\cite{Greco:2003vf,Greco:2003mm}.  Surprisingly, data from the
Relativistic Heavy-Ion Collider (RHIC) for single electrons ($e^{\pm}$)
associated with semileptonic $B$ and $D$ decays in semi-central Au-Au
collisions exhibited a $v_2$ of up to 10$\%$~\cite{Kelly:2004,adare07},
indicating substantial collective behavior of charm ($c$) quarks
consistent with the assumption of a $c$-quark $v_2$ similar to the one
for light quarks, apart from a $p_T$ shift due to radial-flow effects
\cite{Greco:2003vf}. In addition, the nuclear suppression factor was
found to be comparable to the pion one, $R_{AA}$$\simeq$0.3
\cite{Adler:2005xv,abelev07}.  Perturbative QCD (pQCD) calculations of
radiative energy loss cannot explain these findings, even after
inclusion of elastic scattering.  Based on lattice QCD (lQCD) results
which suggest resonance structures in the meson-correlation function at
moderate temperatures, an effective model for heavy-light quark
scattering via $D$ and $B$ resonances was suggested
\cite{vanHees:2004gq}.  When implementing such a picture into a Langevin
simulation the results for semileptonic $e^{\pm}$ spectra
\cite{vanHees:2005wb} are in reasonable agreement with RHIC data
\cite{Adler:2005xv,abelev07,adare07}.

\section{Heavy-Quark Transport in a sQGP Fireball}

We employ a Fokker-Planck approach to evaluate drag and diffusion
coefficients for $c$ and $b$ quarks in the QGP based on elastic
scattering with light quarks via $D$- and $B$-meson resonances with a
width $\Gamma$$=$$400$-$750$~MeV (supplemented by perturbative
interactions)~\cite{vanHees:2004gq}. Heavy-quark (HQ) kinetics in the
QGP is treated as a relativistic Langevin process~\cite{vanHees:2005wb},
and the medium is modeled by a spatially homogeneous elliptic thermal
fireball which expands isentropically.

For RHIC the details of the fireball and the parameters used can be
found in \cite{vanHees:2005wb}; for LHC we extrapolate to $\dd
N_{\mathrm{ch}}/\dd y$$\simeq$1400 for central $\sqrt{s_{NN}}$=5.5~TeV
Pb-Pb collisions.  The expansion parameters are adjusted to hydrodynamic
simulations, resulting in a total lifetime of
$\tau_{\mathrm{fb}}$$\simeq$6~fm/c at the end of a hadron-gas QGP mixed
phase and an inclusive light-quark elliptic flow of $\langle v_2
\rangle$=7.5\%. The main change with respect to RHIC is the expectation
of a new ``phase'' in which hadron-like resonances melt at higher
temperature. In order to implement this ``melting'' of $D$- and
$B$-mesons above $T_{\mathrm{diss}}$=$2\, T_c$$\simeq$360~MeV, a ``transition
factor'' $(1+\exp[(T-T_{\mathrm{diss}})/\Delta])^{-1}$ ($\Delta$=50~MeV)
has been introduced into the transport coefficients.

Initial HQ spectra are computed using PYTHIA with parameters as used by
the ALICE Collaboration~\cite{Alice:2005pp}, and $c$ and $b$ quarks are
hadronized into $D$ and $B$ mesons at $T_c$ by coalescence with light
quarks~\cite{Greco:2003vf}; ``left-over'' heavy quarks are treated by
$\delta$-function fragmentation.
\begin{figure}
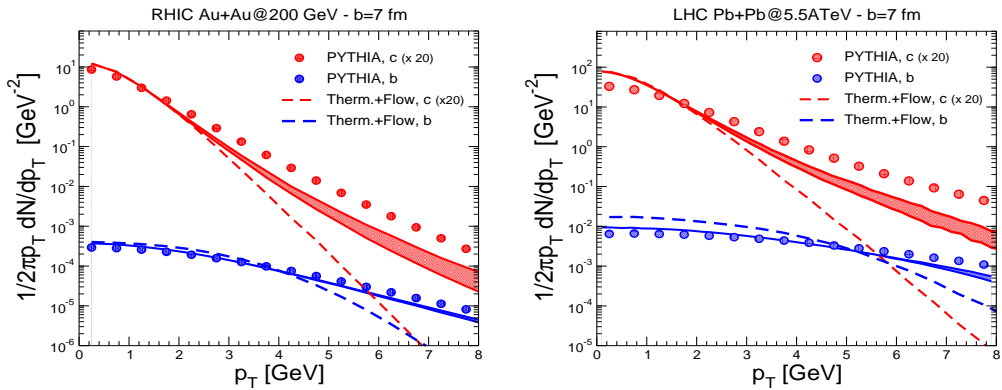

\includegraphics[height=2.in,width=2.5in]{dndpt_bcthermpQCD_RHIC.eps}
\hspace*{0.3cm}
\includegraphics[height=2.in,width=2.5in]{dndpt-HQ-thermpQCD-LHC.eps}
\caption{(Color online) HQ $p_T$ distributions at RHIC (left) and LHC
  (right). Circles are the initial distribution, the bands show the ones
  after interactions in the QGP. Dashed lines are the thermal
  distributions shown for comparison.}
\label{fig1}
\end{figure}

\section{Results and Predictions for RHIC and LHC}

In Fig.~\ref{fig1} the invariant $p_T$ distributions for $c$ (red) and
$b$ (blue) quarks at RHIC (left) and LHC (right) are shown. It is
evident that the charm-quark distributions after rescattering in the
sQGP (band corresponding to the uncertainty in the resonance width) are
close to the thermal distribution for $p_T$$\lesssim$$2m_c$ where most
of the yield resides. On the contrary, $b$ quarks are predicted to be
off-equilibrium (even at LHC energies).

\begin{figure}
\includegraphics[height=1.7in,width=2.5in]{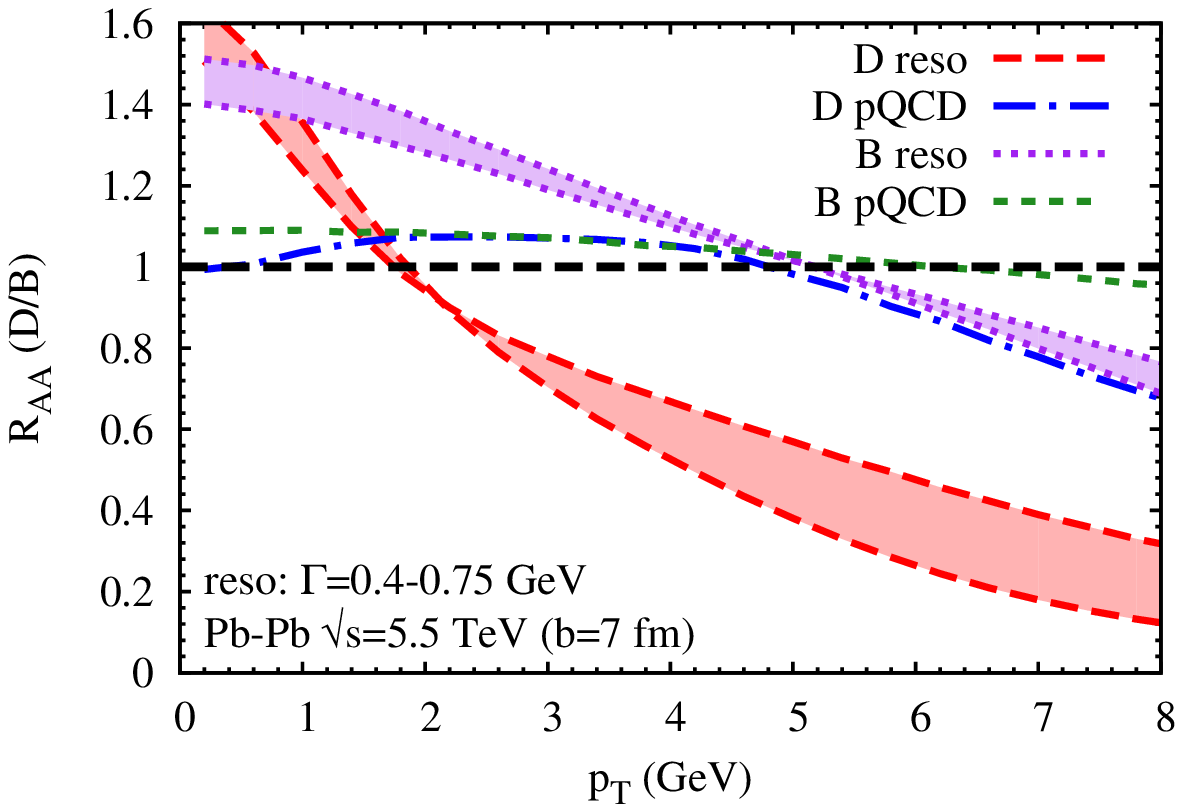}
\hspace*{0.3cm}
\includegraphics[height=1.7in,width=2.5in]{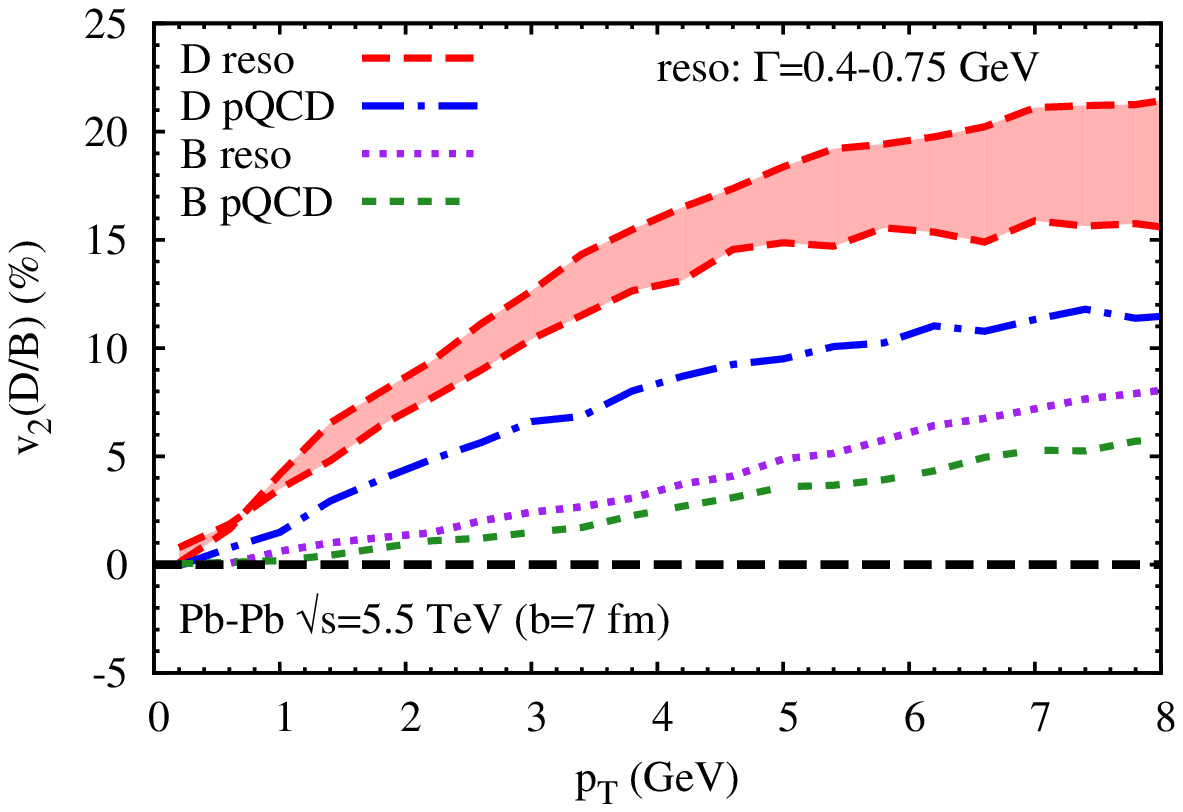}
\caption{(Color online) Predictions of relativistic Langevin simulations
  for heavy qurks in an sQGP for $b$=7~fm, $\sqrt{s_{NN}}$=5.5~TeV
  Pb-Pb collisions: $R_{AA}$ (left) and $v_2$ (right) for $D$ and $B$
  mesons.}
\label{fig2}
\end{figure}

Fig.~\ref{fig2} summarizes our results for HQ diffusion in a QGP in
terms of the meson $R_{AA}(p_T)$ and $v_2(p_T)$ for $b$=7~fm Pb-Pb
collisions at the LHC. Our most important findings are: (a) resonance
interactions substantially increase (decrease) $v_2$ ($R_{AA}$) compared
to perturbative interactions; (b) $b$ quarks are much less affected than
$c$ quarks, reducing the effects in the $e^\pm$ spectra; (c) there is a
strong correlation between a large $v_2$ and a small $R_{AA}$ at the
quark level, which, however, is partially reversed by coalescence
contributions which increase \emph{both} $v_2$ and $R_{AA}$ at the meson
(and $e^\pm$) level. This feature turned out to be important in the
prediction of $e^\pm$ spectra at RHIC; (d) the predictions for LHC are
quantitatively rather similar to our RHIC results~\cite{vanHees:2005wb},
due to a combination of harder initial HQ-$p_T$ spectra (see circles in
Fig.~\ref{fig1}) and a decrease in interaction strength in the early
phases where non-perturbative resonance scattering is inoperative.
Therefore we conclude that if at RHIC the dominant contribution to HQ
interactions are hadron-like resonances, at LHC we should observe an
$R_{AA}$ and $v_2$ pattern similar to RHIC.

\textbf{Acknowledgment:} Work supported by U.S.-NSF under
contract no. PHY-0449489.

\end{document}